\documentclass[sigconf]{acmart}
\usepackage{subfigure}
\usepackage{multirow}
\usepackage{tabularx}
\usepackage{algorithm}
\usepackage{algorithmic}
\usepackage{enumitem}
\usepackage{arydshln}

\usepackage{graphicx}
\usepackage{caption}

\AtBeginDocument{%
  \providecommand\BibTeX{{%
    \normalfont B\kern-0.5em{\scshape i\kern-0.25em b}\kern-0.8em\TeX}}}

\copyrightyear{2023}
\acmYear{2023}
\setcopyright{rightsretained}
\acmConference[SIGIR-AP '23]{Annual International ACM SIGIR Conference on Research and Development in Information Retrieval in the Asia Pacific Region}{November 26--28, 2023}{Beijing, China}
\acmBooktitle{Annual International ACM SIGIR Conference on Research and Development in Information Retrieval in the Asia Pacific Region (SIGIR-AP '23), November 26--28, 2023, Beijing, China}\acmDOI{10.1145/3624918.3625318}
\acmISBN{979-8-4007-0408-6/23/11}

\begin{document}
\definecolor{holden-color-light}{rgb}{0.2, 0.65, 0.99}
\definecolor{holden-color-deep}{rgb}{0.1, 0.3, 0.44}
\newcommand{\holden}[1]{\textcolor{holden-color-light}{#1}}
% \newcommand{\holden}[1]{\textcolor{black}{#1}}

% \title{Frequency-Aware Adversarial Training for Unbiased Factorization Machine}
% \title{Adaptive Adversarial Factorization Machines: Debiased Recommendation for Imbalanced Groups}
\title{Automatic Feature Fairness in Recommendation via Adversaries}

% \author{Anonymous}

\author{Hengchang Hu}
\email{hengchang.hu@u.nus.edu}
\authornote{Corresponding author.}
\affiliation{%
  \institution{National University of Singapore}
  \country{Singapore}
}

\author{Yiming Cao}
\email{caoy0035@e.ntu.edu.sg}
\authornote{Work done while interning at the National University of Singapore.}
\affiliation{%
  \institution{Nanyang Technological University}
  \country{Singapore}
}
% Min: Yiming is now at NTU? I didn't know that. HC: yes

\author{Zhankui He}
\email{zhh004@eng.ucsd.edu}
\affiliation{%
  \institution{UC, San Diego}
  \country{United States}
}

\author{Samson Tan}
\email{samson.tmr@u.nus.edu	}
% Min: please spell-check!
\authornote{Work done while a Ph.D. student at the National University of Singapore.}
\affiliation{%
  \institution{AWS AI Research \& Education}
  \country{United States}
}

\author{Min-Yen	Kan
}
\email{kanmy@comp.nus.edu.sg}
\affiliation{%
  \institution{National University of Singapore}
  \country{Singapore}
}

\renewcommand{\shortauthors}{Hengchang Hu, Yiming Cao, Zhankui He, Samson Tan, \& Min-Yen Kan}

\begin{abstract}
Fairness is a widely discussed topic in recommender systems, but its practical implementation faces challenges in defining sensitive features while maintaining recommendation accuracy.
%and preventing it from compromising recommendation accuracy. 
% In our work, w
We propose \textit{feature fairness} as the foundation to achieve equitable treatment across diverse groups defined by various feature combinations.
This improves overall accuracy through balanced feature generalizability. 
We introduce unbiased feature learning through adversarial training, using adversarial perturbation to enhance feature representation. The adversaries improve model generalization for under-represented features. 
We adapt adversaries automatically based on two forms of feature biases: frequency and combination variety of feature values. This allows us to dynamically adjust perturbation strengths and adversarial training weights. Stronger perturbations are applied to feature values with fewer combination varieties to improve generalization, while higher weights for low-frequency features address training imbalances.
% , resulting in overall improved accuracy.
We leverage the \underline{A}daptive \underline{A}dversarial perturbation based on the widely-applied \underline{F}actorization \underline{M}achine (AAFM) as our backbone model. % which is referred to as AAFM.
In experiments, AAFM surpasses strong baselines in both fairness and accuracy measures.
% , demonstrating superior practicality. 
AAFM excels in providing item- and user-fairness for single- and multi-feature tasks, showcasing their versatility and scalability. To maintain good accuracy, we find that % adversarial perturbation should not persist excessively during training.
adversarial perturbation must be well-managed: during training, 
perturbations should not overly persist and their strengths should decay.

\end{abstract}




\maketitle

\section{Introduction}

%Recommendation systems (RS) are commonly used to address information overload, and fairness in RS has garnered considerable attention.
Fairness in Recommendation Systems (RS) has garnered considerable attention.
Various techniques have been employed, such as outcome re-ranking \cite{singh2018fairness,geyik2019fairness,li2021user} and unbiased learning \cite{ai2018unbiased,geyik2019fairness,yu2020influence,zhang2021causal,sato2020unbiased} (mitigating biases in the training process directly). 
However, the specific fairness requirements vary depending on the stakeholders and the specific needs of the application. 
The definition of fairness in user-centric or item-centric recommendations relies on the chosen sensitive features \cite{wu2022selective}.
Prior studies \cite{li2021user,geyik2019fairness,zhu2021fairness,li2021tutorial} only consider the chosen sensitive features either from users or items, which poses challenges in terms of fairness scalability when considering the other aspect. 
Furthermore, imposing constraints to achieve fairness often compromises overall recommendation accuracy, further constraining real-world applicability.

\begin{figure}[t] 
\centering
    \setlength{\abovecaptionskip}{-0.05cm}
    \setlength{\belowcaptionskip}{-0.3cm}
  \subfigure{
		\label{fig:intro_case}
		\includegraphics[width=0.49\linewidth]{intro_case.pdf}}
    \subfigure{
		\label{fig:m_home_stu_plo}
		\includegraphics[width=0.43\linewidth]{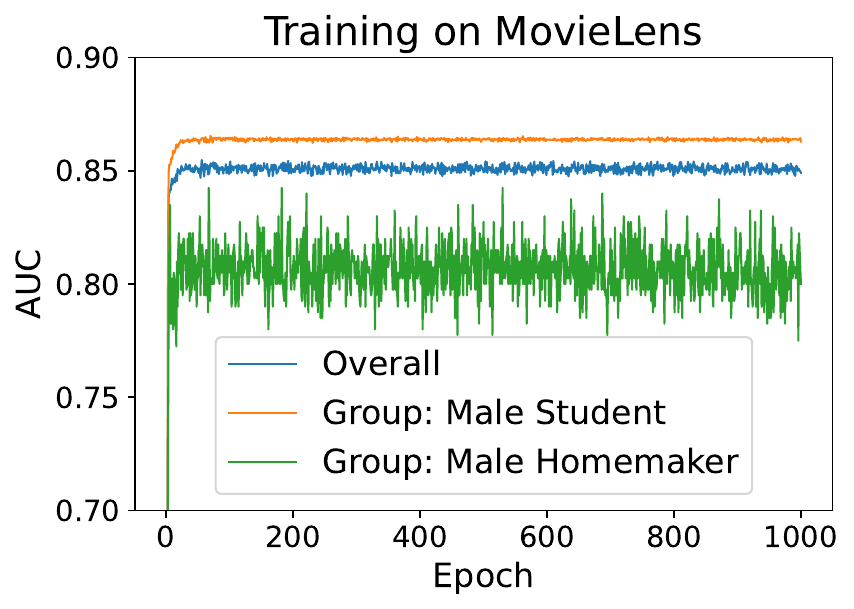}}
    \caption{(left) Unfairness between two groups with sensitive features. (right) Biased validation accuracy between two data groups during training of Factorization Machine.}
	\label{fig:intro}
\end{figure}

To enable flexible selection of sensitive features, we introduce generic \textbf{feature fairness} as our core guiding principle.
It centers on features themselves, agnostic to whether features are from users or items.
In this work, we examine two statistical biases for feature values (e.g., \textit {student} or \textit{male}): feature \textit{frequency}, indicating the occurrence rate within its feature domain (e.g., user occupation, or user gender); and feature \textit{combination variety}, representing the diversity of co-occurring samples with other features. 

To investigate the biased outcomes resulting from skewed features, let us take a case of MovieLens. 
Our preliminary analysis focuses on two feature-defined user groups: \textit{male+students} and \textit{male+homemakers}. 
We use Factorization Machine \cite{rendle2010factorization} for modeling here.
Figure~1 illustrates that the majority group of \textit{male students} (representing 21.09\% of users) consistently outperforms the average, while the minority group of \textit{male homemakers} (representing 0.1\%) exhibits below-average performance with significant fluctuations during training. This disparity in accuracy and stability results in unfairness. There are two causes: (1) Limited co-occurrence frequency of the \textit{male} and \textit{homemaker} features in training hinders the model's ability to capture their interactions. 
(2) Additionally, when there is a greater combination variety of gender for the \textit{student} feature compared to the \textit{homemaker} feature, the model struggles to recognize interactions between the \textit{homemaker} feature and different gender values, resulting in poorer generalization.

% Based above observation, in this work, we examine two statistical biases for feature values (e.g., student): feature \textit{frequency}, indicating the occurrence rate within its feature domain (e.g., user occupation); and feature \textit{combination variety}, representing the diversity of co-occurring samples with other features. 
In this work, we aim to utilize the two forms of feature biases to automatically (1) incorporate fairness considerations across diverse feature domains; and (2) ensure similar generalizability for different combinations of feature values.

Adversarial training \cite{goodfellow2014explaining} is a technique for augmenting model generalization \cite{he2018adversarial}, where the generalization derives from its robustness to unseen inputs. 
We thus adopt adversarial training to accommodate a variety of feature combinations.
By integrating adversarial training into our regular training iterations, we enhance feature representations by perturbing them. However, applying this approach directly still poses issues.

First, existing approaches assume consistent perturbation intensities \cite{he2018adversarial,chen2019adversarial} for all feature representations, but there are significant variations in sample outcomes associated with different features. 
Our method utilizes \textit{combination variety} as the measure to determine the intensity of adversarial perturbation. We employ a formula that maps lower variety values to higher adversarial intensity, thereby enhancing the stability of targeted groups. To prevent excessive perturbation that overly distorts the original data representation, we map \textit{variety} inversely proportional to a range of $0\sim1$.
Second, conventional adversarial unbiased learning approaches often view accuracy and fairness as conflicting objectives \cite{wu2022selective}. 
As features often follow a long-tailed distribution, low-frequency features make up the majority of features. Hence, low-frequency features are important, so we prioritize their appropriate representation during training by assigning higher adversarial training weights. This balancing results in enhanced performance.

We instantiate the above-mentioned Adaptive Adversaries with the FM model as our backbone, or AAFM for short. 
Extensive experiments show that our method improves results by 1.9\% in accuracy against baselines, while balancing group standard deviation by $\frac{7}{10}$ on fairness metrics.
AAFM further demonstrates scalability, tackling fairness concerns for both users and items simultaneously. 
Additionally, as the number of feature domains in the data increases, our approach consistently tackles fairness at finer levels among diverse groups. This serves as a bridge between group and individual fairness, spanning datasets with one feature domain to those with a broader range of three feature domains. 
% Min: still redundant
Our method's universal applicability to fairness issues % should not be underestimated. Importantly, it 
offers a win--win outcome by promoting both fairness and accuracy.

In summary, our contributions are as follows: 
(i) Compared to user fairness and item fairness, we define our task as a more fundamental feature fairness objective. The feature fairness task aims to develop a parameter-efficient framework that flexibly provides feature-specific fairness for various combinations of user or item features.
(ii) We introduce AAFM, an adversarial training method that leverages statistical feature bias for unbiased learning, combining the benefits of fairness and accuracy.
(iii) Through experiment datasets with varying numbers of features, user- and item-centric settings, we validate the scalability and practicality of AAFM in real scenarios.
The code is available at: \url{https://github.com/HoldenHu/AdvFM}

\section{Methodology}
\label{sec:approach}

In what follows, we first outline our task and delve into the issue of feature fairness, which arises due to two biases.
We then provide our solution --- Adversarial Factorization Machines which applies the fast gradient method to construct perturbations over feature representations. We further propose an adaptive perturbation based on feature biases, which re-scales adversarial perturbation strengths and adversarial training weights.

\subsection{Preliminaries of Feature Fairness}
% \noindent \textbf{Task of Feature Fairness.}
% In real-world applications, the selection of sensitive features (from users or items) and their combinations can vary depending on the demands.
% Our task objective is to \textbf{(1)} ensure that different features within the same domain have similar recommendation accuracy for their respective data subsets. For instance, in terms of user fairness, the prediction accuracy should be comparable between all data associated with students and homemakers within the occupation feature domain.
% \textbf{(2)} Different combinations of features should also exhibit similar generalizability. For example, the presence of a larger variety of co-occurring occupations for males compared to females as a feature combination creates unfairness for the female feature. This results in poorer generalization during the model inference stage, as the model fails to recognize the interactions between the female feature and a broader range of occupational types.
% Compared to user fairness and item fairness, we define our task as a more fundamental feature fairness objective. The feature fairness task aims to develop a parameter-efficient framework that can flexibly provide feature-specific fairness for various combinations of user or item features.

\textbf{Problem Formulation.}
The recommendation task aims to predict the probability of unobserved user--item interactions $\hat{y}(\mathbf{x})$ given the user and item features $\mathbf{x}$ \cite{hu2023adaptive}. We represent one sample, the input as the combination of these features, denoted as $ \mathbf{x} = \{x_1, x_2, ..., x_n\}$.
Here, $x_i$ represents the $i^{th}$ feature domain, encompassing user features (e.g., user occupation) and item features (e.g., item color). 
Concerning the $k^{th}$ sample $\mathbf{x}^{(k)} \in \mathcal{X}$, $x_i^{(k)}$ indicates its specific feature value (e.g., \textit{student} or \textit{red}) in feature domain $x_i$.
In our work, the feature domains include user/item ID, and the categorical attributes of user/item.
Concerning specific feature value $v$ in domain $x_i$, we denote its corresponding samples of subset data as $\mathcal{X}_{x_i:v} = \{\mathbf{x}^{(k)} | x_i^{(k)} = v \}$. The overall prediction error of the subset data is denoted as $\mathcal{E}_{x_i:v} = \sum_{\mathbf{x} \in \mathcal{X}_{x_i:v}} \mathcal{E}(\hat{y}(\mathbf{x}), y)$. Here, $\mathcal{E}$ indicates the metric (e.g., Logloss) measuring errors between the prediction $\hat{y}$ and ground-truth $y$, where $y \in \{0,1\}$.

To achieve feature fairness, we expect a smaller difference between errors $\mathcal{E}_{x_i: v_1}$ and $\mathcal{E}_{x_i: v_2}$ with respect to each feature domain $x_i$ and each value pair $(v_1, v_2)$ within $x_i$. 
In neural models, the precise representation of each value is vital, as it directly affects errors in corresponding samples. The quality of feature value representation depends on the statistical bias (e.g., popularity bias \cite{ran2022pm}) of feature values in the data.

\textbf{Two Forms of Feature Biases.}
Feature values in the data distribution have the following statistical properties. To aid understanding, we show an example of feature value $v$ in the feature domain $x_i$.
\begin{itemize}[leftmargin=*]
    \item \textit{Frequency} $\alpha_{v}$ indicates the occurrence rate of the value $v$ concerning its feature domain.
    \item \textit{Combination variety} $\beta_{v}$ indicates the number of diverse samples where value $v$ co-occurs with other features in combination.
\end{itemize}
$\alpha_v$ can be used to measure how many times this feature value has been seen by the model, while $\beta_v$ better reflects the degree of isolation of this feature-based data group. The more isolated the groups are, the more likely they are sensitive to model perturbation.

\begin{figure}[t!]
    \setlength{\abovecaptionskip}{-0cm}
    \setlength{\belowcaptionskip}{-0.5cm}
    \centering
    \includegraphics[width=0.48\textwidth]{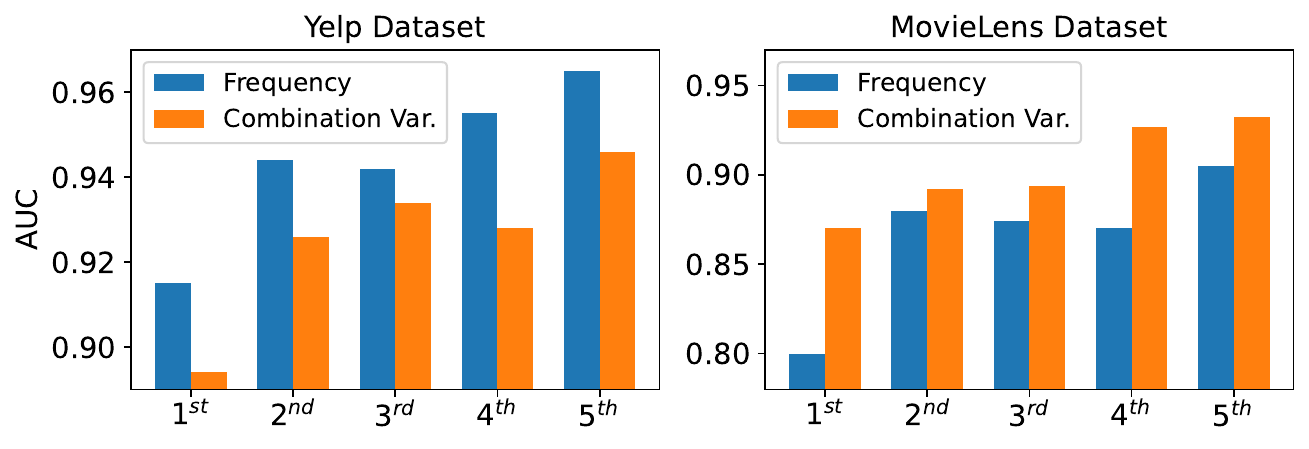}
    \caption{Unbalanced results regarding two forms of feature biases. x-axis indicates the indices of sample groups sorted by the overall feature frequency/combination variety. The results are from FM applied to the Yelp/Movielens dataset.}
    \label{fig:framework}
\end{figure}

In normal distributions, combination variety and frequency can be viewed as equivalent, where the frequency increase, the combination variety increase as well.
But in real-world cases, this may not hold true as feature values may not always follow a strict joint probability dependence. 
Take the feature domain gender as an example. Given a situation where \textit{female} has fewer combinations with \textit{occupation} than \textit{male}, this does not mean that the feature value \textit{female}'s frequency is necessarily less than \textit{male}. 
In the results depicted in Figure~2, data samples were grouped into 5 bins based on the multiplied value of frequency or combination variety across all feature domains (\textit{user features+item features}). While both biases contribute significantly to performance imbalances, they are not aligned, highlighting the interdependence between features in real-world data.
Therefore, we consider them as separate statistical biases for utilization.

\subsection{Adversarial Factorization Machine (AdvFM)}

\subsubsection{Base Model}
Our framework consists of three stages (Figure ~), characterized by stages for Embedding. Representation learning and Prediction.

\paragraph{(a) Embedding Initialization}
To improve the representative ability of features, we first map each original discrete feature value of $x_i$ into $d$-dimensional continuous vectors $e_i = \mathcal{M}(x_i | \Theta)$ through the embedding layer $\mathcal{M}$. Here, the concatenated feature embeddings are denoted as $\mathbf{e}=cat[e_1; ...;e_n]$.

\paragraph{(b) Representation Learning}
Our key insight is that the inter-dependencies among low-level feature groups play a critical role in robustness and fairness.  For this reason, we use Factorization Machines (FM)~\cite{rendle2010factorization} as the backbone for our methodology.  FM takes a set of vector inputs, each consisting of $n$ feature values and performs recommendations through their cross-product.  An FM model of degree $2$ estimates the rating behavior $\hat{y}$ as:  
\begin{equation}
    f(\mathbf{e}) = 
    \sum_{i=1}^{n} \left\langle w_i, e_i \right\rangle
    +\sum_{i=1}^{n} \sum_{j=i+1}^{n}\left\langle v_{i}, v_{j} \right\rangle e_{i} e_{j} ,
\end{equation}
\noindent where parameter $w_i \in \mathcal{R}^{1 \times d}$ models the linear, first-order interactions, and $v_{i} \in \mathbb{R}^{1 \times d}$ models second-order interactions for each low dimensional vector $e_i$. 
$\left\langle\cdot , \cdot \right\rangle$ indicates the dot product operation and $e_{i} e_{j}$ indicates element-wise product between them.
To be concise, we use the notation $\hat{y}(\mathbf{x}|\Theta) = f(\mathbf{e})$ to represent the model's processing of input $\mathbf{x}$ with the embedding parameter $\Theta$.

\paragraph{(c) Prediction \& Model Training}
The training objective function is defined as:
\begin{equation}
    \mathcal{L}(\hat{y},y) = \sum_{(\mathbf{x}, y)} y \log(\hat{y}(\mathbf{x}|\Theta) + (1-y)\log(1-\hat{y}(\mathbf{x}|\Theta))
\end{equation}
where $\mathcal{L}$ indicates the cross entropy loss \cite{good1952rational}, the difference between the predicted and true values.

\begin{figure}[t!]
    \setlength{\belowcaptionskip}{-0.5cm}
    \centering
    \includegraphics[width=0.48\textwidth]{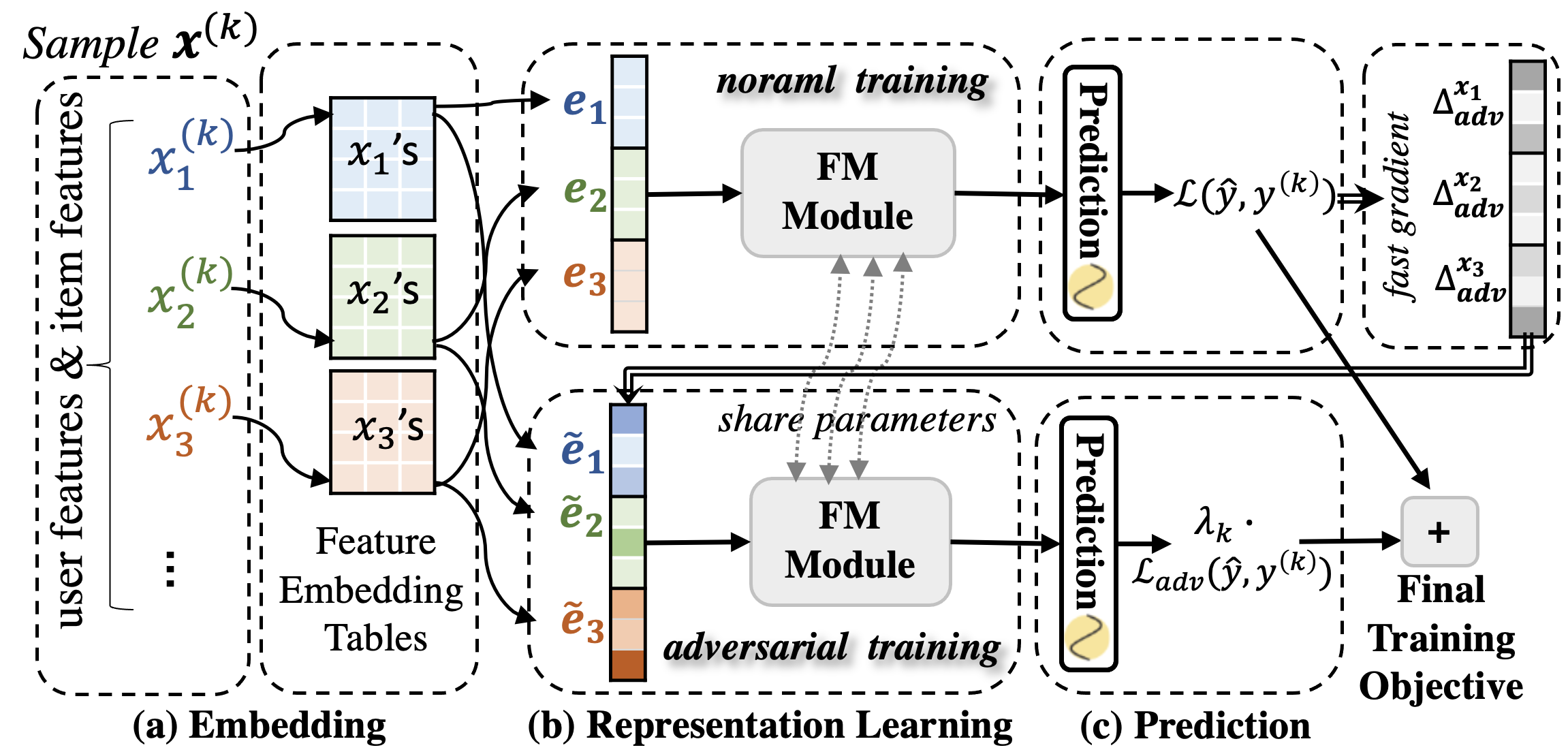}
    \caption{The training process of adversarial factorization machine on sample $\mathbf{x}^{(k)}$. }
    \label{fig:framework}
\end{figure}

\subsubsection{Adversarial Perturbation}
Inspired by previous work \cite{shivaswamy2022adversary} which observed that users with rare interactions would benefit more from robustness, we adopt gradient-based adversarial noise \cite{goodfellow2014explaining} as the perturbation mechanism to improve balanced robustness from the feature perspective.

As shown in Fig~\ref{fig:framework}, the normal representation learning of FM module utilizes the original embedding $\mathbf{e}$.
The adversarial training adds noise to each feature's embedding by perturbing FM's parameters:
\begin{equation}
    \tilde{e}_i = \mathcal{M}(x_i|\Theta+\Delta_{adv}^{e_i})
\end{equation}
where $\Delta_{adv}^{e_i}$ is the parameter noise providing the maximum perturbation on the embedding layer.
$\Delta_{adv} = \{\Delta_{adv}^{e_1},... \Delta_{adv}^{e_n}\}$ denotes the overall perturbations on embedding layer.

To efficiently perturb normal training, we estimate the optimal adversarial perturbation by maximizing the loss incurred during training:
\begin{equation}
    \Delta_{adv}=
    {\arg \max_{\|\delta\| \leq \epsilon} } 
    \mathcal{L}(\hat{y}( \mathbf{x} |\Theta+\delta), y),
\label{eqn_delta}
\end{equation}
where the hyper-parameter $\epsilon$ controls the strength level of perturbations, and $\|\cdot\|$ denotes the $l2$ norm. 
% Min: please check citation placement % HC: Done
Our adversarial noise uses the backward propagated fast gradient \cite{goodfellow2014explaining} of each feature's embedding parameters as their most effective perturbing direction.
Specifically, to perturb the embedding $e_i$, we calculate the partial derivative of the normal training loss:
\begin{equation}
    \Delta_{adv}^{e_i}=
    \epsilon \cdot 
    \frac{\partial \mathcal{L}(\hat{y}(\mathbf{e}|\Theta), y) / \partial e_i}
    {\| \partial \mathcal{L}(\hat{y}(\mathbf{e}|\Theta), y) / \partial e_i \|} ,
\end{equation}
% Min: please check the clarification % HC: Done
where the right-hand side's normalized term is the sign of the fast gradient direction of the feature $x_i$'s embedding parameters.

\paragraph{Training objective}
In each epoch, we conduct training as normal first, then introduce the adversarial perturbations in another following training session, round by round. We define the final optimization objective for AdvFM as a min--max game:
\begin{equation}
    {\arg \min_{\Theta} }  \{ { \arg \max_{\Delta_{adv}} } 
    [ \mathcal{L}(\hat{y}(\mathbf{x}|\Theta), y) + \lambda \cdot \mathcal{L} (\hat{y}(\mathbf{x}|\Theta+\Delta_{adv}), y) ]\}
\label{eq:goal} 
\end{equation}
% , where $\Delta_{adv}$ is the overall noise providing the maximum perturbation on the embedding layer,
where $\Delta_{adv}$ provides the maximum perturbation and $\Theta$ is trained to provide a robust defense to minimize the overall loss. Here, $\lambda$ is a hyper-parameter to control the adversarial training weights. 

\subsection{Automatic Adaptation on AdvFM}
\label{sec:aafm}

% Min: check below. HC: Done
The approach described so far has a key drawback: 
It introduces a single, uniform perturbation strength level $\epsilon$ overall features, and uniform adversary weights $\lambda$ over all samples.  This makes the method inflexible, and unable to model nuanced weighting. 
To further balance and improve the accuracy, we further propose an Adaptive version of AdvFM (AAFM).
It auto-strengthens the adversarial perturbations on the feature embedding parameters, and re-weights the samples in adversarial training. Our adaptive version leverages the two forms of feature biases previously introduced (Fig~\ref{fig:framework}, right). 

\begin{itemize}[leftmargin=*]
    \item \textit{Auto-Strengthening.} 
    % Min: check wording HC: Reject
    Considering each feature domain $x_i$ with the corresponding value $v_i$, a smaller combination variety $\beta_{v_i}$ indicates a higher degree of sensitivity representation.
    Thus, it needs to be trained with stronger perturbations on its embedding parameters to improve its robustness. We estimate the feature-specific $\epsilon_{v_i}$ based on an inversely proportional basis:
    \begin{equation}
    \epsilon_{v_i}= \psi \left( \omega_{i} \times (\beta_{v_i})^{-1}\right) ,
    \end{equation}
    where $\omega_{i}$ is a learnable parameter with respect to the feature domain $x_i$. We adopt SoftPlus activation function for $\psi$, as it does not change the sign of the gradient, and the SoftPlus unit has a non-zero gradient over all real inputs. 
    \item \textit{Re-Weighting.} Unlike previous work \cite{he2018adversarial} conducting fixed adversarial training weight $\lambda$ for all samples, we conduct sample-specific ones.
    Specifically, given a sample $\mathbf{x}^{(k)}$, the sample-specific adversary weight $\lambda_{k}$ is defined as:
    \begin{equation}
    \lambda_{k}= \Phi ( -\prod_{x \in \mathbf{x}^{(k)}} \alpha_x , t) ,
    \end{equation}
    For the sample $\mathbf{x}^{(k)}$ with a low overall feature frequency $\prod_{x \in \mathbf{x}^{(k)}} \alpha_{x}$ in training, we increase the weight of its adversarial loss by increasing its associated $\lambda$ value.
    The function $\Phi(\cdot, t)$ is used to scale the values between 1 and $t$. If we use the previous design of trainable parameter $\omega$ to scale, $\lambda$ is easily eliminated by the overarching optimization goal (Equation~\ref{eq:goal}); hence we apply manually-controlled  scaling via $t$.
\end{itemize}

\paragraph{Optimization of Decaying Adversarial Perturbation}

When the model adaptively adjusts the adversarial perturbation (noise) level $\epsilon$, we observe that optimization may simply set $\epsilon$ to zero, which best meets the normal training objectives by achieving a local optimum.  However, this thwarts the benefit of introducing adversarial perturbation; canceling it prematurely. 

To mitigate this, we envision a slow decline in the effect of adversarial perturbation, proportional to the time already trained.

To this end, we design a regulation term for $\omega$ by defining an additional loss $\mathcal{L}_{decay}=\alpha ({\tau} \cdot { \| \omega \| })^{-1}$, where  $\tau$ represents the trained epoch number, and $\alpha$ is an annealing hyper-parameter controlling regulation strength.
As such, the change of $\omega$ is more marked during early training, where a small $\omega$ would make $\mathcal{L}_{decay}$ large. As the training proceeds and the model stabilizes, the sensitivity of $\omega$  gradually decays, as $\tau$ increases.

\section{Experiments}
\label{sec:exp}

\textbf{Datasets.}
We experiment on three public datasets to examine our model's debiasing effect on both user and item groups. 
User feature enriched recommendation datasets include movie dataset \textit{MovieLens-100K}\footnote{https://grouplens.org/datasets/movielens/} (user gender, occupation, and zip code), and image dataset \textit{Pinterest}\footnote{https://sites.google.com/site/xueatalphabeta/academic-projects} (user preference categories). 
Item feature enriched recommendation datasets include movie dataset \textit{MovieLens-100K}$^{1}$ (movie category, and release timestamp), and business dataset \textit{Yelp}\footnote{https://www.yelp.com/dataset} (business city, star).
% BUG: Min: filtered out or only retained...? HC: The writing is right, filtering out. Done
Following the previous work \cite{he2017neural} to reduce the excessive cost, we filtered out the user with more than 20 interactions in Yelp, and randomly selected 6,000 users to construct our Pinterest dataset.
We convert all continuous feature values into categorical values (e.g., by binning user age into appropriate brackets), and consider the user and item IDs as additional features.
% as additional categorical features as well

\vspace{0.4em}

\noindent \textbf{Baselines.}
% As we are not only concerned about our model's recommendation accuracy but also the debias effects,  we select state-of-the-art methods for both discussions.
% \textit{Accuracy Baselines} include matrix factorization-based method ONCF \cite{he2018outer} and FM-families --- benchmark FM \cite{rendle2010factorization}, NFM \cite{he2017neural}, DeepFM  \cite{guo2017deepfm}, CFM \cite{xin2019cfm}.
% \textit{Debiasing Baselines} include regularization-based approach M-Match \cite{kamishima2013efficiency}, classical inverse propensity scoring approach IPS \cite{saito2020unbiased}, MACR \cite{wei2021model}  incorporating user/item's effect in the loss, DecRS \cite{wang2021deconfounded} investigating the causal representation of users.
We choose our comparison baseline with respect to models achieving strong recommendation accuracy and debias effects.
\textit{Accuracy Baselines} include matrix factorization-based method ONCF \cite{he2018outer} and FM-families --- FM \cite{rendle2010factorization}, NFM \cite{he2017neural}, DeepFM  \cite{guo2017deepfm}, CFM \cite{xin2019cfm}.
\textit{Debiasing Baselines} include regularization-based approach M-Match \cite{kamishima2013efficiency}, classical inverse propensity scoring approach IPS \cite{saito2020unbiased}, MACR \cite{wei2021model}  incorporating user/item's effect in the loss, DecRS \cite{wang2021deconfounded} investigating the causal representation of users.

\vspace{0.4em}

\noindent \textbf{Evaluation Protocols.}
% For the train--test data splitting, we employ the standard \textit{leave-one-out} strategy \cite{bayer2017generic,he2018adversarial}.
% For evaluating the \textit{accuracy} of our model, we use the commonly adopted AUC (Area under the ROC Curve) and Logloss (cross-entropy). 
% To further study the group \textit{fairness} concerning imbalanced features, we propose two quantitative metrics.
% We first rank the data samples $\mathbf{x}$ concerning their feature frequency and co-occurrence, i.e., $\prod_{x \in \mathbf{x}} (Fr_x \cdot Co_x)$. \holden{Confusing sentence.}. Then, we quantify the unfairness based on the ranked samples.
% (i) EFGD (Extreme Feature-based Groups Difference): 
% Following the previous practice \cite{pedreschi2009measuring} that term the difference between the two extreme data groups as the indicator, We take the AUC difference between the first 10\% samples and the last 10\%  in our case.
% (ii) STD (Standard Deviation): 
% To measure more fine-grained fairness as \cite{wei2021model}, we divide the ranked samples into 6 bins. The AUC standard deviation of the 6 bins is another measurement of the fairness degree.
% \holden{Re-check the citation}
For the train--test data split, we employ standard leave-one-out \cite{he2018adversarial}.
To evaluate the \textit{accuracy}, we adopt \textit{AUC} (Area Under Curve) and \textit{Logloss} (cross-entropy). 
To assess the \textit{fairness} concerning imbalanced features, we split data into buckets for evaluation, following previous work \cite{pedreschi2009measuring,wei2021model}.
We first rank the data samples $\mathbf{x}^{(k)}$ by joint feature statistics $\prod_{x \in \mathbf{x}^{(k)}} (\alpha_x \cdot \beta_x)$, and divide the ranked samples into 6 buckets.
We propose two quantitative metrics as follows. 
\begin{itemize}[leftmargin=*]
    \item \textit{EFGD} (extreme feature-based groups difference). Following the previous practice \cite{pedreschi2009measuring} that term the difference between the two extreme data groups as the indicator, we take EFGD as the AUC difference between the first 10\% samples and the last 10\%.
    \item \textit{STD} (overall groups' standard deviation). STD is used to measure more fine-grained fairness (as \cite{wei2021model}). And STD stands for the AUC standard deviation of the buckets.
\end{itemize} 

\begin{table*}[t!]
\centering
% \scriptsize
% \tabcolsep=1.15mm
\renewcommand{\arraystretch}{1.2}
\begin{tabular}{c|cc|ccccc|ccccc}
\hline
\hline
\textit{Scenarios} & \textit{Dataset} & \textit{Metrics} & FM & NFM & CFM & DeepFM & ONCF & AdvFM & AAFM$^\lambda$ & AAFM$^\epsilon$ & AAFM  & D-AAFM  \\
\hline
\multirow{4}{*}{\textbf{\textit{item-centric}}} & \multirow{2}{*}{ML$^i$} & LL & 0.4093 & 0.3688 & 0.3635 & 0.3597 & 0.3641 & 0.3730 & 0.3391  & 0.3673 & 0.3352 & \textbf{0.3248}  \\
 & & AUC & 0.9154 & 0.9203 & 0.9257 & 0.9381 & 0.9243 & 0.9337 & 0.9406  & 0.9308 & 0.9408 & \textbf{0.9431}  \\
\cline{2-13}
& \multirow{2}{*}{Yelp} & LL   & 0.1934 & 0.1895 & 0.0963 & 0.1584 & 0.1527 & 0.1692 & 0.0878  & 0.1751 & \textbf{0.0731}  & 0.0742 \\
& & AUC & 0.9474 & 0.9569 & 0.9732 & 0.9665 & 0.9668 & 0.9653 & 0.9790  & 0.9619 & \textbf{0.9813}  & 0.9795 \\
\hline
\multirow{4}{*}{\textbf{\textit{user-centric}}} & \multirow{2}{*}{ML$^u$} & LL   & 0.4493 & 0.4297 & 0.3876 & 0.3109 & 0.3721 & 0.4325 & 0.3182  & 0.4323 & 0.3072 & \textbf{0.2996}  \\
 & & AUC & 0.8796 & 0.8908 & 0.9172 & 0.9319 & 0.9012 & 0.8810 & 0.9249  & 0.8808 & 0.9323 & \textbf{0.9357}  \\
\cline{2-13}
& \multirow{2}{*}{Pinterest} & LL   & 0.5647 & 0.3865 & 0.3577 & 0.3541 & 0.4026 & 0.3859 & 0.3573  & 0.3914 & 0.3447 & \textbf{0.3042}  \\
 & & AUC & 0.5700 & 0.7430 & 0.7356 & 0.7580 & 0.7251 & 0.7432 & 0.7695  & 0.7408 & 0.7756 & \textbf{0.8031} \\
\hline
\hline
\end{tabular}
\caption{Overall accuracy performance comparison. Smaller LL (Logloss) or larger AUC indicates better accuracy. ML$^{i}$ or ML$^{u}$ indicate the partial MovieLens dataset with only item or item features.
AAFM$^\lambda$ and AAFM$^\epsilon$ only adaptively adjust $\lambda$ (with fixed $\epsilon=0.5$) and $\epsilon$ (with fixed $\lambda=1$) respectively. D-AAFM indicates AAFM incorporating decaying perturbation regularization.}
\label{tab:overall-performance}
\end{table*}

\begin{table*}[t!]
\setlength{\abovecaptionskip}{-0cm}
\setlength{\belowcaptionskip}{-0cm}
\centering
% \scriptsize
% \tabcolsep=0.45mm
\renewcommand{\arraystretch}{1.3}
\begin{tabular}{c|cc|ccccc|ccccc}
\hline
\hline
\textit{Scenarios} & \textit{Dataset} & \textit{Metrics} & FM   & IPS   & M-match & MACR  & DecRS & AdvFM & AAFM$^{\lambda}$ & AAFM$^{\epsilon}$ & AAFM       & D-AAFM     \\
\hline
\multirow{4}{*}{\textbf{\textit{item-centric}}} & \multirow{2}{*}{ML$^i$}  & EFGD & 0.0713 & 0.0336 & 0.0381 & 0.0282 & 0.0589 & 0.0401 & 0.0232      & 0.0241      & 0.0110      & \textbf{0.0105} \\
& & STD & 0.0257 & 0.0237 & 0.0236 & 0.0171 & 0.0246 & 0.0235 & 0.0139      & 0.0161      & 0.0091      & \textbf{0.0069} \\
% \hdashline
\cline{2-13}
& \multirow{2}{*}{Yelp}   & EFGD & 0.0440 & 0.0243 & 0.0301 & 0.0177 & 0.0272 & 0.0230 & 0.0166      & 0.0228      & \textbf{0.0144} & 0.0181     \\
& & STD & 0.0131 & 0.0114 & 0.0153 & 0.0082 & 0.0122 & 0.0086 & 0.0082      & 0.0079      & \textbf{0.0064} & 0.0068     \\
\hline
\multirow{4}{*}{\textbf{\textit{user-centric}}} & \multirow{2}{*}{ML$^u$}  & EFGD & 0.0415 & 0.0289 & 0.0337 & 0.0294 & 0.0374 & 0.0340 & 0.0323      & 0.0377      & \textbf{0.0281} & 0.0368     \\
& & STD & 0.0280 & 0.0198 & 0.0208 & 0.0199 & 0.0230 & 0.0225 & 0.0219      & 0.0259      & \textbf{0.0195} & 0.0220     \\
% \hdashline
\cline{2-13}
& \multirow{2}{*}{Pin.} & EFGD & 0.1068 & 0.0682 & 0.0726 & 0.0558 & 0.0545 & 0.0853 & 0.0289      & 0.0580      & 0.0193     & \textbf{0.0132} \\
& & STD & 0.0307 & 0.0277 & 0.0299 & 0.0275 & 0.0265 & 0.0300 & 0.0213      & 0.0296      & 0.0195      & \textbf{0.0178} \\
\hline \hline
\end{tabular}
\caption{Feature fairness effect comparison. 
The smaller the STD or EFGD, the fairer the results.
The abbreviations are the same as in Table~\ref{tab:overall-performance}. The upper/lower two datasets correspond to item-centric/user-centric fairness.}
\label{debias_performance}
\end{table*}

\subsection{Recommendation Accuracy Comparison}
\label{sec:accuracy}

\subsubsection{Superior Accuracy Against Baselines}
We present the overall results in Table~\ref{tab:overall-performance}.
Regarding both user and item feature-enriched datasets, our AAFM consistently outperforms other FM-based baselines.
Among the baselines, DeepFM achieves the best performance in three datasets, as indicated by both Logloss and AUC metrics. This highlights its effectiveness in mapping sparse features to dense vectors using the neural embedding layer.
CFM, employing 3D CNN, outperforms ONCF, which uses 2D CNN, indicating the superiority of 3D CNN in extracting feature interactions.

\subsubsection{Ablation Study}
To further investigate where the performance improvement of AAFM originates from, we present the ablation study in the right-hand columns of Table~\ref{tab:overall-performance}.
We can see that compared to AdvFM (without any adaptive optimization), the introduction of adaptive $\lambda$ significantly enhances the overall performance. 
% Min: check phrasing and sentence reordering HC: Done
This indicates that our proposed adversarial training reweighting is promising and can optimize well, instead of locking the fairness model within performance-compromising constraints.

However, introducing only adaptive $\epsilon$ worsens the overall performance on several datasets. By considering both aspects together, synthesizing them into AAFM, and adding decaying perturbation regularization loss, we get D-AAFM. Either of them performs best across all datasets. In most cases, D-AAFM performs better, demonstrating that persistent adversarial perturbations can severely impact model accuracy.

\subsection{Feature Fairness Results}
\label{sec:debias}

\subsubsection{Superior Fairness Against Baselines}
Feature fairness is another aspect of concern in our study. As depicted in Table~\ref{debias_performance}, all fairness baselines show improvement over FM in terms of metrics measuring the reduction in bias (EFGD and STD). 
We observe that the phenomenon of feature unfairness does exist, and that current fairness models do alleviate this issue. Among the baselines, MACR performs the best; it considers the popularity bias of both users and items, taking into account the impact of skewed occurrences of user or item IDs.
Our AdvFM also provides more fair results, compared to FM. However, it is not as good as the aforementioned debiasing baselines. 
This corroborates that though adversarial training has shown promise in promoting fairness recently, it necessitates further detailed investigation.
Through careful design of adversarial perturbations, our AAFM and D-AAFM achieve better fairness, concerning either user features or item features.

\subsubsection{Ablation Study}
To figure out how the effects of adversaries improve fairness, we conduct an additional ablation study, shown in the right columns of Table~\ref{debias_performance}.
Compared to AdvFM, the inclusion of adaptive $\lambda$ and adaptive $\epsilon$ both significantly contribute to improving fairness.
When both are utilized (i.e., AAFM), the effect on feature fairness is further enhanced. This demonstrates that both proposed automatic adaptations are complementary and indispensable. Features with smaller combination variety require a larger $\epsilon$ to improve generalization ability. Even though we encourage it by using the reciprocal of its bias, it is still very easy to reduce $\epsilon$ during training (thereby reverting back to normal training). In order to forcefully encourage adversarial training, it is necessary for samples with less frequent features to have more adversarial training weight, thus enabling the adversaries to truly play their role.
Similar to the finding from the accuracy comparison, D-AAFM and AAFM alternately become the best models, suggesting different dataset sensitivities to long-term perturbations.

\begin{figure*}[t]
    \centering
    \includegraphics[width=0.9\textwidth]{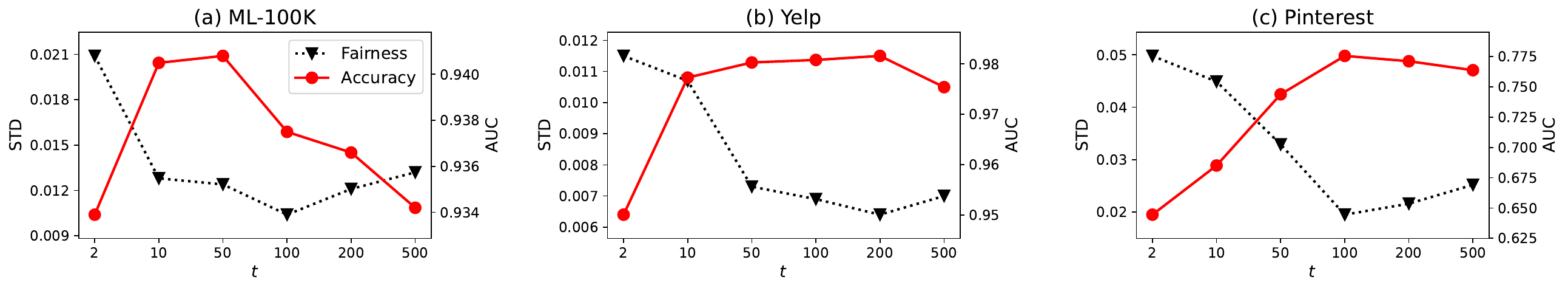}
    \caption{Trade-off between accuracy and user group fairness via control of the re-weighting parameter $t$. Smaller STD ($\downarrow$) indicates better fairness, and larger AUC ($\uparrow$) indicates better accuracy.}
    \label{fig:tradeoff}
\end{figure*}

\subsection{Robustness of AdvFM} 

\begin{table}[t]
\small
\tabcolsep=1.2mm
\centering
\begin{tabular}{l|ccc|ccc|ccc}
\hline
\textit{Dataset}   & \multicolumn{3}{|c}{\textbf{ML-100K}} & \multicolumn{3}{|c}{\textbf{Yelp}}    & \multicolumn{3}{|c}{\textbf{Pinterest}} \\
\textit{Noise} & 0.5     & 1.0     & 2.0     & 0.5     & 1.0     & 2.0     & 0.5      & 1.0      & 2.0     \\
\hline
FM  & -4.67 & -9.58 & -18.9 & -6.14 & -12.7 & -24.3 & -2.74  & -3.40  & -4.95 \\
AdvFM    & -2.32 & -4.75 & -9.60 & -3.38 & -6.79 & -13.3 & -1.48  & -1.53  & -1.64 \\
AAFM     & -0.64 & -0.76 & -1.00 & -1.41 & -3.03 & -6.37 & -0.29  & -0.30  & -0.32 \\
\hline
\end{tabular}
\captionof{table}{Performance drop ratio (\%) in AUC of models in the presence of external adversarial perturbation.}
\label{tab:robust}
% \vspace{-0.8cm}
\end{table}

Driven by the premise that adversarial training enhances robustness for perturbed parameters, we delve into understanding this improvement. 
In order to probe the robustness of groups under feature representation perturbations, we adopt the methodology from \cite{he2018adversarial}, which infuses external noise into the model parameters at levels spanning 0.5 to 2.0.
As shown in Table~\ref{tab:robust}, we observe that AdvFM exhibits less sensitivity to adversarial perturbations compared to FM. For instance, on the Yelp dataset, a noise level of 0.5 results in a decrease of $6.14\%$ for FM, whereas AdvFM only experiences a decrease of $3.38\%$. Moreover, AAFM demonstrates even greater stability with a decrease of only $1.41\%$. 
From the perspective of these improvements in robustness, we see the model's ability to generalize to unseen inputs, giving indicative evidence for why rare features are handled well by our proposed methods.

\textit{Case Study.}
The benefits of such robustness improvement are particularly pronounced for small groups characterized by less frequent features and unstable performance during training. To illustrate this, we select the \textit{male entertainment} group, which accounts for only 0.2\% of the total users, as a case study (Figure~4). The figure demonstrates that normal FM training exhibits significant fluctuations, indicating the sensitivity of the data group to model updates. In contrast by incorporating annealing adaptive noise in AAFM, performance gradually converges while improving overall AUC in the later stages of training. This notable improvement in stability further confirms the enhanced robustness in small groups.

\begin{figure}
\vspace{-5pt}
  \begin{minipage}{.29\textwidth}
    \centering
    \includegraphics[width=\linewidth]{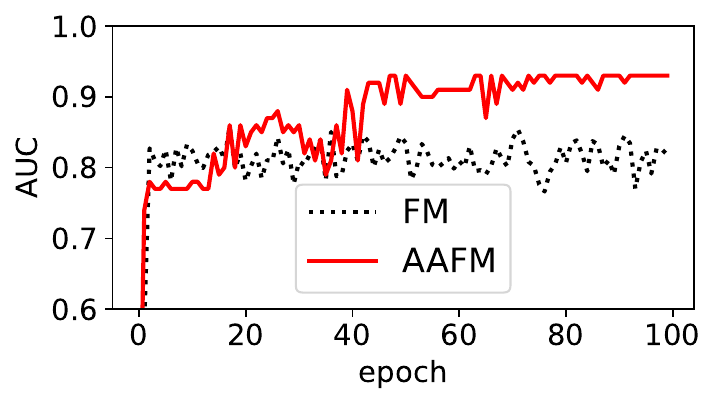}
  \end{minipage}
  \begin{minipage}{.18\textwidth}
    \vspace{-20pt}
    \captionof{figure}{Validation accuracy (AUC) on the small group \textit{(male entertainment)} in MovieLens dataset. The case study.}
  \end{minipage}
\vspace{-15pt}
\end{figure}

\subsection{Trade-off Between Fairness and Accuracy} 

Fairness and accuracy often involve a trade-off, and sometimes their objectives can even be contradictory \cite{wei2021model}. However, we argue that fairness and accuracy can find common ground with appropriate adaptive adversarial weights.
We adjust the hyperparameter $t$ to control the scale of $\lambda_k$ in Figure~\ref{fig:tradeoff}. As $t$ increases, we observe that fairness achieves the best results when $t$ takes on the values of 100, 200, and 100 for MovieLens, Yelp, and Pinterest, respectively. On the other hand, accuracy reaches its peak when $t$ is set to 50, 200, and 100. Notably, these two objectives are mostly aligned, suggesting that the improvement in fairness mainly stems from the enhanced accuracy of small groups rather than compromising the performance of larger groups (which could significantly reduce overall accuracy). The exception occurs in the MovieLens dataset, where there is a trade-off between the best accuracy ($t=50$) and the best fairness $t=100$.
MovieLens contains more feature domains compared to the other two datasets. This implies a finer feature granularity and more similar joint feature statistics for samples.
Larger $t$ will magnify the differences in adversarial weights of samples that were originally similar.
This will lead to a rapidly increasing amount of samples with low training weights, resulting in a more prominent overall performance drop.

\section{Related Work}
\label{sec:rw}

% AAFM relates to both fairness and adversarial methods in recommendation, and as such we review both areas in the following.

\textbf{Fairness in recommendation} is a nascent but growing topic of interest~\cite{beutel2019fairness}, but hardly has a single, unique definition. The concept has been extended to cover multiple stakeholders\cite{abdollahpouri2020connection,singh2018fairness} and implies different trade-offs in utility. 
From a stakeholder perspective, fairness can be considered from both item and user aspects. \textit{User fairness}~\cite{li2021user,geyik2019fairness} expects equal recommendation quality for individual users or demographic groups, and \textit{item fairness} \cite{zhu2021fairness,li2021tutorial} indicates fair exposure among specific items or item groups. 
From an architectural perspective, there are mainly two approaches to address fairness: 
One method is to post-process model predictions (i.e., re-ranking) to alleviate unfairness~\cite{singh2018fairness,geyik2019fairness,li2021user}.
The other unbiased learning method is to directly debias in the training process.
Such latter methods come from two origins.
Causal Embedding~\cite{bonner2018causal} is one way to control the embedding learning from the bias-free uniform data (e.g., by re-sampling~\cite{geyik2019fairness}).
Re-weighting \cite{yu2020influence,burke2017balanced} is another method to balance the impact of unevenly distributed data during training, where the Inverse Propensity Scoring \cite{zhang2021causal,sato2020unbiased} is a common means to measure the difference between actual and expected distributions. 
In this work, we generalize the problem to solve both user and item groups' unfairness, proposing an unbiased learning technique at the feature-level.

\vspace{0.4em}

\noindent \textbf{Adversarial training in recommendation} helps models pursue robustness by introducing adversarial samples. 
One of the most effective techniques is to perturb adversarial samples by gradient-based noise (e.g., FGSM~\cite{goodfellow2014explaining}, PGD~\cite{madry2017towards}, and C\&W~\cite{carlini2017towards}).
Previous work found such noise is effective in improving recommendation accuracy, such as applying fixed FGSM on matrix factorization \cite{he2018adversarial} and multiple adversarial strengths \cite{yuan2019adversarial-FG}.
Current adversarial perturbation in recommendation systems mostly focuses on representing individual users \cite{beigi2020privacy,li2021towards,shivaswamy2022adversary} or items \cite{beigi2020privacy,krishnan2018adversarial} properly.

Adversarial training is increasingly discussed in unbiased learning approaches \cite{wu2021fairrec}.
Recent work \cite{shivaswamy2022adversary} also found adversarial perturbation could benefit under-served users. 
\citet{yu2022graph} found a positive correlation between the node representation uniformity and the debias ability, and added adversarial noise to each node in contrastive graph learning.
However, they lack systematic comparison with fair recommendation baselines and overlook the flexibility of selected features. While there have been discussions in computer vision on connecting fairness and model robustness \cite{xu2021robust,wei2022prototypical}, there is a lack of studies addressing the bridging between model robustness and the co-improvement of accuracy and fairness in recommendation tasks.

\section{Conclusion and Future Work}
\label{sec:con}
%\vspace{-0.2em}

% Compared to user fairness and item fairness, we define our task as a more fundamental feature fairness objective. The feature fairness task aims to develop a parameter-efficient framework that can flexibly provide feature-specific fairness for various combinations of user or item
% features

%% Conclusion
In this work, we propose a feature-oriented fairness approach, employing feature-unbiased learning for simultaneous improvement of fairness and accuracy. We address imbalanced performance among feature-based groups by identifying its root causes in feature frequency and combination variety. Our proposed Adaptive Adversarial Factorization Machine (AAFM) uses adversarial perturbation to mitigate this imbalance during training, applying varied perturbation levels to different features and adversarial training weights to different samples. This adaptive approach effectively enhances the generalizability of feature representation. Our experimental results show that AAFM outperforms in fairness, accuracy, and robustness, highlighting its potential as an effective approach for further study in this field.

While AAFM introduces adversarial training to unbiased learning, there are still many possible refinements.  For example, AAFM defaults to using random negative sampling, which biases toward the majority of users/items features.
How to balance the impact of such biased negative sampling in different groups deserves future study.
It will also be valuable to further investigate the effectiveness of different adversaries (e.g., PGD~\cite{madry2017towards}, or C\&W~\cite{carlini2017towards}) on more complex neural recommendation backbones.

% \begin{acks}
% none
% \end{acks}

\appendix
% \subfile{sec/appendix.tex}
\section{Derivation of Adversarial Perturbation}

% \subsection{Derivation of Adversarial Perturbation }
We present the mathematical derivation of the adversarial perturbation for feature embedding $e_{i}$, and explain the reasoning behind utilizing combination variety as the bias parameter to achieve balance.

By applying the Chain Rule, we express the adversarial feature perturbation $\Delta_{adv}^{e_i}$ in the following manner:
\begin{equation}
\begin{aligned}
    \Delta_{adv}^{x_i} & =
    \epsilon \cdot 
    \frac{\partial \mathcal{L}(\hat{y}, y) / \partial{e_i}}
    {\| \partial \mathcal{L}(\hat{y}, y) / \partial{e_i} \|} \\
    & = 
    \epsilon \cdot 
    \frac{\left(-\frac{y}{\hat{y}}-\frac{1-y}{1-\hat{y}}\right) \cdot  {\partial \hat{y}} / {\partial e_i}}
    {\|\left(-\frac{y}{\hat{y}}-\frac{1-y}{1-\hat{y}}\right) \cdot {\partial \hat{y}} / {\partial e_i} \|} \\
\end{aligned}
\end{equation}
$y$ can take on values of either 0 or 1, hence we can simplify the above expression as:
\begin{equation}
    \Delta_{adv}^{e_i} = -
    \epsilon \cdot 
    \frac{ {\partial \hat{y}} / {\partial e_i}}
    {\|\ {\partial \hat{y}} / {\partial e_i}\|} 
\end{equation}

\noindent Given that we have chosen FM as our prediction model, we can calculate the partial derivative of $\hat{y}$ with respect to the feature embedding $e_i$ as follows:

\begin{equation}
\begin{aligned}
% \hat{y} & =\langle w, e\rangle+\sum_{i=1}^n \sum_{j=i+1}^n\left\langle v_i, v_j\right\rangle e_i e_j \\
\frac{\partial \hat{y}}{\partial e_i} & =w_i+\frac{\partial}{\partial e_i}\left[\frac{1}{2} \sum_{f=1}^d\left(\sum_{i=1}^n v_{i, f}  e_i\right)^2-\frac{1}{2} \sum_{f=1}^d\left(\sum_{i=1}^n v_{i, f}^2 e_i^2\right)\right] \\
& =w_i+\frac{1}{2}\left[\frac{\partial}{\partial e_i} \sum_{f=1}^d\left(v_{i , f}^2 e_i^2+2 \sum_{j=1}^n v_{i, f} v_{j, f} e_i e_j\right)-\frac{\partial}{\partial e_i} \sum_{f=1}^d\left(\sum_{i=1}^n v_{i, f}^2 e_i^2\right)\right] \\
% & =w_i+\frac{1}{2}\left[\frac{\partial}{\partial e_i} \sum_{f=1}^d\left(v_i \cdot f e_i^2\right)+2 \frac{\partial}{\partial e_i} \sum_{f=1}^d\left(\sum_{j=1}^n v_{i \cdot f} v_{j, f} e_i e_j\right)-\frac{\partial}{\partial e_i} \sum_{f=1}^d\left(\sum_{i=1}^n v_{i, f}^2 e_i^2\right)\right] \\
& =w_i+\sum_{f=1}^d\sum_{\substack{j=1 \\ j \neq i}}^n v_{i, f} v_{j, f} e_j \\
& =w_i+\sum_{\substack{j=1 \\ j \neq i}}^n<v_i, v_j> e_j
\end{aligned}
\label{ap_eq: ye}
\end{equation}

\noindent where $\frac{\partial}{\partial e_i} \sum_{f=1}^d\left(\sum_{i=1}^n v_{i, f}^2 e_i^2\right)$ can be reduced and vector multiplication involved is performed element-wise. Substituting ${\partial \hat{y}} / {\partial e_i}$ into $\Delta_{adv}^{e_i}$, we thus have:
\begin{equation}
% \tilde{e_i}=e_i+
\Delta_{adv}^{e_i}=-\epsilon \cdot \frac{w_i+\sum_{\substack{j=i \\ j \neq i}}^n\left\langle v_i, v_j\right\rangle e_j}{\left\|w_i+\sum_{\substack{j=1 \\ j \neq i}}^n\left\langle v_i, v_j\right\rangle e_j\right\|}
\end{equation}

The addition of this adversarial perturbation to the original embedding $e_{i}$ utilizes the interacted feature embeddings $e_j$ weighted by the pair-wise interaction weights $\left\langle v_i, v_j\right\rangle$ to enhance the representation of embedding $e_{i}$.

Hence, we can find the perturbation on $e_i$ is controlled by the strength $\epsilon$, and the perturbation direction is influenced by $w_i$ and $v$. There exists a direct relationship between $w_i$ and the perturbation direction. As for $\left\langle v_i, v_j\right\rangle$, being the second-order interaction parameters, their pairwise combinations determine the impact of other $e_j$ values on the perturbation direction.
When $w_i$ is held constant, a larger variety of feature combinations results in a more diverse range of perturbation directions. Consequently, in our work, we assign a smaller perturbation strength to balance between the influence of adversaries and their impact.

\bibliographystyle{ACM-Reference-Format}
\bibliography{sample-base}

\end{document}